\documentstyle[multicol,prl,aps,epsf,psfig,epsfig]{revtex}

\begin{document}

\title
{Modelling Aging Characteristics in Citation Networks}
\author
{Kamalika Basu Hajra and Parongama Sen}
\address
{
Department of Physics, University of Calcutta,
    92 Acharya Prafulla Chandra Road, Kolkata 700009, India. \\
}

\maketitle
\begin{abstract}

Growing network models with
 preferential attachment 
dependent on  both age and degree are proposed to simulate certain features
of citation network noted in \cite{red2}. In this directed network,
a new node gets attached to an older node
with the probability $\sim K(k)f(t)$ where 
the  degree and age of the older node are $k$ and  
 $t$  respectively.
Several  functional forms of 
$K(k)$ and $f(t)$   have been considered. The desirable features 
of the citation network can be reproduced  with $K(k) \sim k^{-\beta}$ 
and $f(t) \sim \exp(\alpha t)$ with $\beta =2.0$ and $\alpha = -0.2$ and with
simple modifications in the growth scheme.

\end{abstract}

Preprint no: CU-Physics-14(2005)


\begin{multicols}{2}

\section{ Introduction}

The citation patterns of scientific publications form a rather complex network.
Here the nodes are published papers and a link is formed if one paper cites
another paper published previously. In  \cite{red1} the citation distribution of $783,339$
papers cataloged by Institute of Scientific Information (ISI)
 and also the $24,296$ papers published in Physical Review D (PRD) between 
$1975$ and $1994$ was studied. It was found that the probability $P(k)$ that a particular paper is 
cited $k$ times follows a power law distribution $P(k)\sim k^{-{\gamma}}$
with exponent $\gamma =3$,
indicating that the incoming degree distribution of the citation network
is scale-free.
Later these studies were extended  \cite{vazquez}  to the outgoing degree
 distributions as well, and it was shown that it has an exponential tail in most cases.\\
 
   The citation distribution provides an interesting platform for
theoretical modelling when the various features of citation dynamics are taken 
into account. It must be kept in mind that citation is possible only to papers 
that have been published {\it previously}, i.e, {\it older} papers, so that the
network is  {\it directed}. Also since 
most of the papers are gradually forgotten or become irrelevant, the probability that a particular
 paper is cited should decrease in time unless it is of utmost importance.
  Again, a {\it young} paper, which is undergoing recognition, gains increasing
attention through citations. Hence the model of a citation network should be
 one in which {\it aging} of the papers occur such that the 
 probability of a paper
getting cited  depends on its age.
  Again, from the scale-free nature of the degree
distribution, it appears that  the probability of a paper 
being cited at a given time is proportional to 
its in-degree. 

The distribution of ages of cited papers was studied for small sample 
sizes in \cite{zhu} and \cite{kamo1.ps}
and the results from these
two studies did not agree.
    The complete set of citations for all 
publications in Physical Review (PR) journals from July $1893$ to June $2003$
was  later studied in  \cite{red2} which perhaps gives the
closest possible picture of the citation scenario.

Among the various features of
a citation network, those which   
  are relevant to the present paper are listed below:\\
(i) the distribution $T(t)$ of  ages $t$ of citations to other 
publications:  
 this is calculated  from the difference of the year of
publication of a particular paper  and the year of publication
of the papers which are cited by it.\\
(ii) the distribution $R(t)$ of citation ages $t$ from citing publications
 calculated  from the difference of the year of
publication of a particular paper  and the year of publication
of the papers citing it.

Fig. 1 shows pictorially how the two distributions are generated.
\begin{figure}
\includegraphics[clip,width =7cm, angle=0]{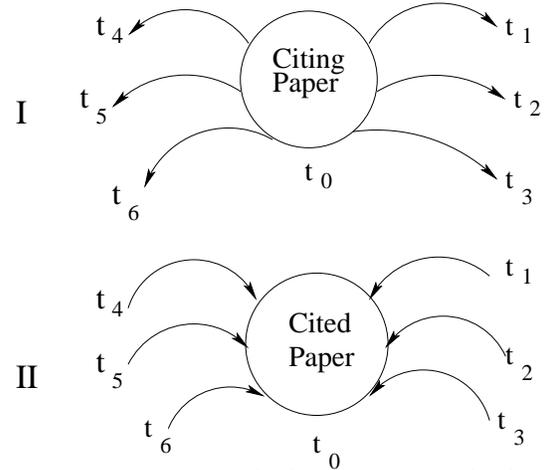}
\caption{The two age distributions from   a citation network. 
In I, the paper published  at
time $t_0$ cites several papers published at different times 
$t_1, t_2$ etc. 
The distribution of the intervals $(t_0-t_i)$
gives $T(t)$.
 In II, the paper published at time $t_0$ is cited by papers published at 
times $t_1, t_2$ etc. 
The distribution of the intervals $(t_i-t_0)$
gives $R(t)$.}
\end{figure}

(iii) The correlation of the average age of citing papers as a function
of  the
degree $k$ of that paper: this is denoted by $A(k)$. 
It is expected that for a paper with many citations
the average age of the citations will also be large such that
there is a positive correlation between the two. 

 For $T(t)$  it was found that in the range of $2$ to $15$ years, the distribution 
decays {\it exponentially} with time, while for longer times the decay is a 
slower exponential. For $R(t)$, over the limited range of $2$ to $20$ years,
the integrated data is consistent with a power law decay with an exponent
$\sim -1$. Hence, authors tend to have an exponentially decaying memory
while citing papers, but the  citation age distribution to a 
particular paper
has a slower power law decay over an initial period of
time ({\it lifetime} of the paper).     
 The PR data  showed that there is indeed  a positive correlation between
average citation age and the number of times that a paper has been cited
(property (iii)) and the relation is consistent with a power law.

In the present paper we have attempted to obtain a suitable model for the 
citation network such that it  may reproduce some of the main results
 that were obtained
from the study of  real citation networks. In section II, we give
 a brief review of time dependent networks, where we discuss the results
of some earlier works.
 In section III,
the results for $R(t)$ from the known models are discussed and we find that these models
 are not appropriate for the citation network.
 In section IV, we propose a  modified model which can reproduce
some of the real results to an appreciable extent.
 Finally in section V, we provide a summary and 
also the conclusions of the present work.\\
 
\section{Brief Review of Models of Aging  Networks}

   The question of time dependence in the attachment probability of the 
incoming nodes in a growing network has been studied in a few theoretical 
models \cite{zhu,DM,kamo2.ps}. These models have basically
evolved from the original Barabasi-Albert (BA) model \cite{BArev} where in a growing
network model, 
a new node gets linked to the existing ones 
 following a preferential attachment to nodes with larger degree.
In the time dependent  models, a new node gets attached
 to older nodes with a preferential attachment which is  dependent on the degree as well
as the age of the existing node. We discuss briefly below some 
relevant age dependent models and the results thereof.\\
  
In  general, in all the models of aging networks,
the attachment probability $\Pi(k,t)$ is  taken to be a separable 
function of the degree $k$
 and the age  $t$ of the existing node such that 
\begin{equation}
{\Pi}(k,t)= K(k) f(t).
\end{equation}
  In the Dorogovtsev-Mendes (DM) model \cite {DM}, 
$K(k) = k$ and $f(t) = t^\alpha$ were considered.

In this model the degree distribution was found to be scale free  
 for values of $\alpha \ge -1$. For $\alpha < 0$, 
the age dependence presents a competing effect to the preferential attachment, 
but for $\alpha > 0$, the older nodes get richer, enhancing the
 '{\it rich gets richer}' effect.\\

In  \cite{zhu} an  exponential decaying function
$f(t) = \exp(\alpha t)$ was chosen and it was found that the model is not scale-free
for any negative value of $\alpha$.

 In \cite{kamo2.ps}, the DM model was further generalised by incorporating
 a power law variation of the  degree in the attachment probability $\Pi$,

\begin{equation}
\Pi(k,t) \sim k^{\beta}t^{\alpha}.
\end{equation}
A phase diagram was obtained for this model in the 
$\alpha - \beta$ plane, with the phase 
boundary dividing the phase space into the  small world
  and regular network regions. Scale free
 behaviour was found to exist  only along a line for 
$\beta \ge 1$. In the small
 world region, there was gel formation beyond $\beta = 1$, while the degree 
distribution was stretched exponential for $\beta < 1, \alpha \le -1$.\\

\section{R(t) from standard models}

Evidently  a time dependent model would  be appropriate 
for the citation network.
One can immediately realise that the time dependent part $f(t)$  of 
the preferential attachment probability (1) is 
analogous to the function
$T(t)$ defined in section I.
 The task  is to investigate  
whether assuming an exponential decay in  $T(t)$ (i.e., $(f(t)$) 
gives us the proper 
behaviour of $R(t)$.

In our theoretical model, we first  take  two  standard forms of 
time dependence in $\Pi(k,t)$  and look at the
behaviour of the corresponding $R(t)$ using a numerical simulation.
 The decay of $f(t)$ is assumed to be  (a) power law  and (b) 
exponential. 
 The  choice 
of a power law behaviour  in the attachment probability 
may be regarded as  of theoretical interest mainly
as $T(t)$ has been observed to have an exponential decay \cite{explain}.
However, the power law model is quite well studied and it may be useful to 
get the results from both models and compare them with the real data.
We also use a power law dependence of $K(k)$ on $k$.

The degree distribution has already been studied for most of these
models. Therefore we are primarily interested in calculating $R(t)$, 
which is related to the degree distribution when its average is 
under consideration. 

 In our simulations we have generated networks with $2000$ nodes and $10000$ 
configurations for the power law  time dependence of the attachment
 probability, while for the exponential
time dependence, we have used a maximum of $3000$ nodes and $5000$
 configurations.
 
Let the $i$th node born at time $\tau_i$ get $R(\tau,\tau_i)$ links at time 
$\tau$. We are interested in the behaviour of  $R(\tau,\tau_i)$ as a function
of the corresponding age $\tau-\tau_i=t$. 
It may be noted that the cumulative sum 
\begin{equation}
 R_{cum}(\tau,\tau_i) = \Sigma_{\tau'=\tau_i}^{\tau} R(\tau', \tau_i)
\end{equation}
is a well-studied quantity in many networks as a function of $\tau$ and  
$\tau_i$ and in many network models  like the BA or DM model it behaves as 
\begin{equation}
 R_{cum}(\tau,\tau_i) = {\cal {R}}(\tau / \tau_i)
\end{equation}
where ${\cal {R}}(x)$ has a power law growth for large $x$, e.g., ${\cal {R}}(x)
\propto x^{1-\rho}$ ($\rho < 1$). In more complicated models, e.g., 
accelerated models \cite{ps_acc}, 
$ R_{cum}(\tau,\tau_i)$ may have  a non-trivial 
dependence on both $\tau$ and $\tau_i$.
In any case, as a function of 
$t$, $R_{cum}$ will have a strong $\tau_i$ dependence.  
For the distribution of the ages of citing papers,    
we therefore find it more convenient to tag an arbitrary  node
  and study  the number of links $R(t)$ it gets
as a function of $t$   
suppressing the index $\tau_i$. 
The price we pay for this is that since there is no averaging there
is greater fluctuation. The node we tag also has to be an
early one such that data over a long period is obtainable.

In the following, we detail our findings from the simulations
using two different schemes.

III.I {\it Scheme(a)}: In the first scheme, the attachment probability 
is given by  
$\Pi(k,t) \sim k^{\beta}t^{\alpha}$.
We have simulated the network for $\beta=0.5,1.0$ and $2.0$ and
different values of $\alpha \leq 0$. Throughout the simulations, we have tagged
node number 10 (the results do not change if we change this number 
keeping it an early node). The $\beta=1$ case corresponds to the DM model.
From the behaviour of $R_{cum}(\tau,10)$ here, one can guess that
$R(t)$ will have a form
\begin{equation}
R(t) \propto (\frac {t+10}{10})^{-\rho}.
\end{equation}
This behaviour is observed for large values of $t$ and
the agreement becomes worse as $\alpha$ becomes more negative.
We are more interested in the small $t$ behaviour here, which
turn out to be far from a power law.

\begin{center}
\begin{figure}
\includegraphics[clip,width= 5cm, angle=270]{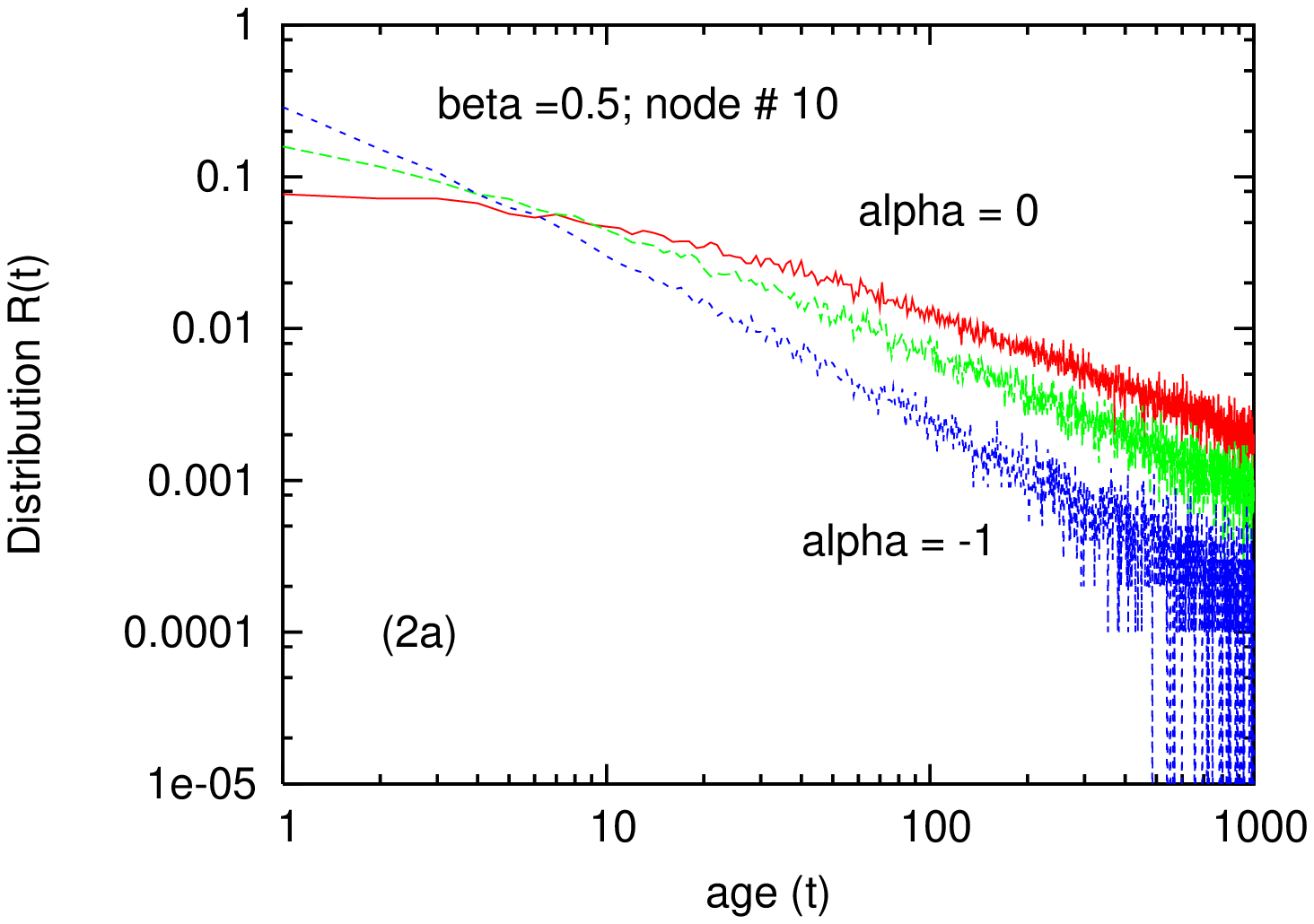}
\includegraphics[clip,width= 5cm, angle=270]{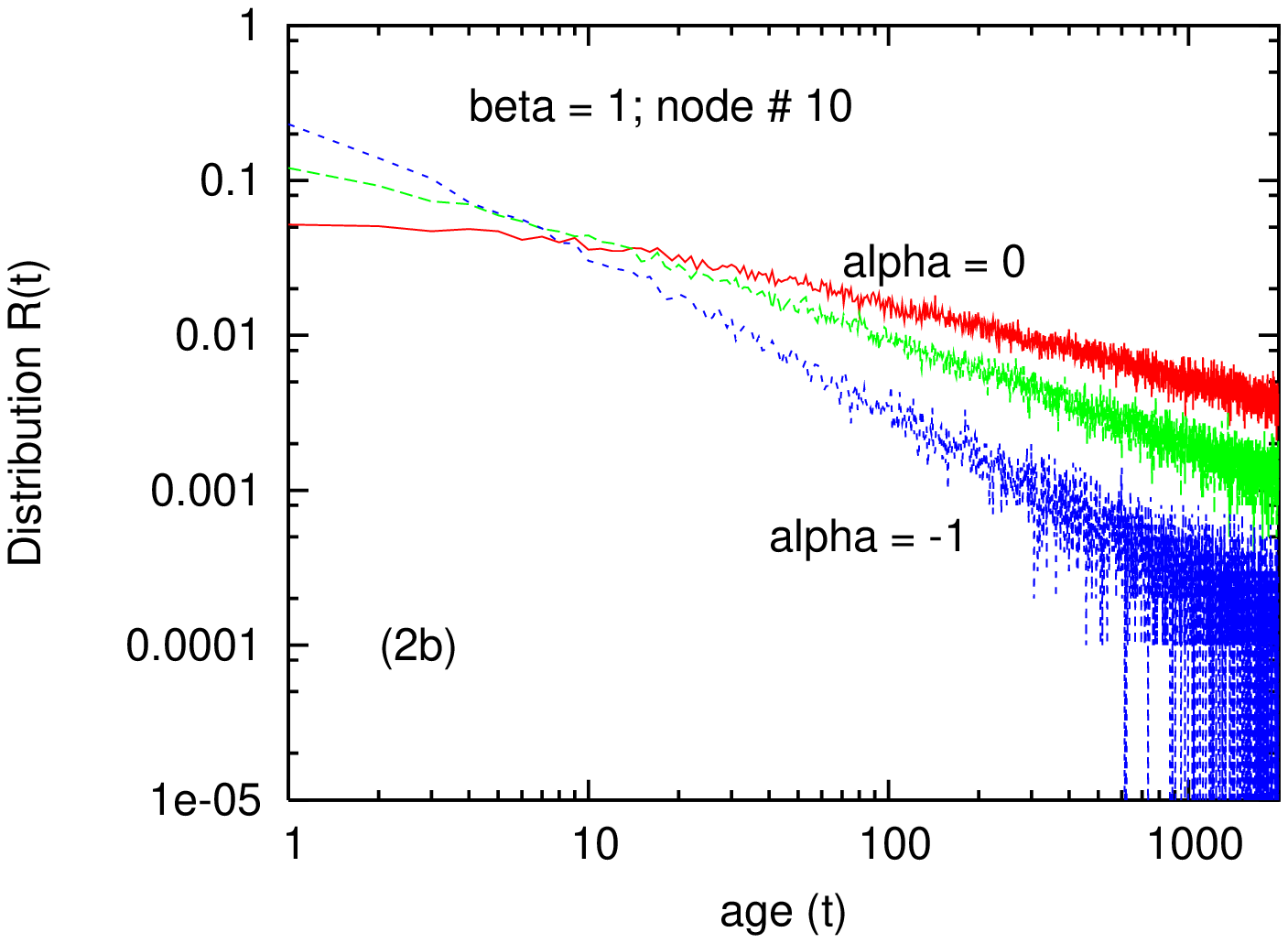}
\includegraphics[clip,width= 5cm, angle=270]{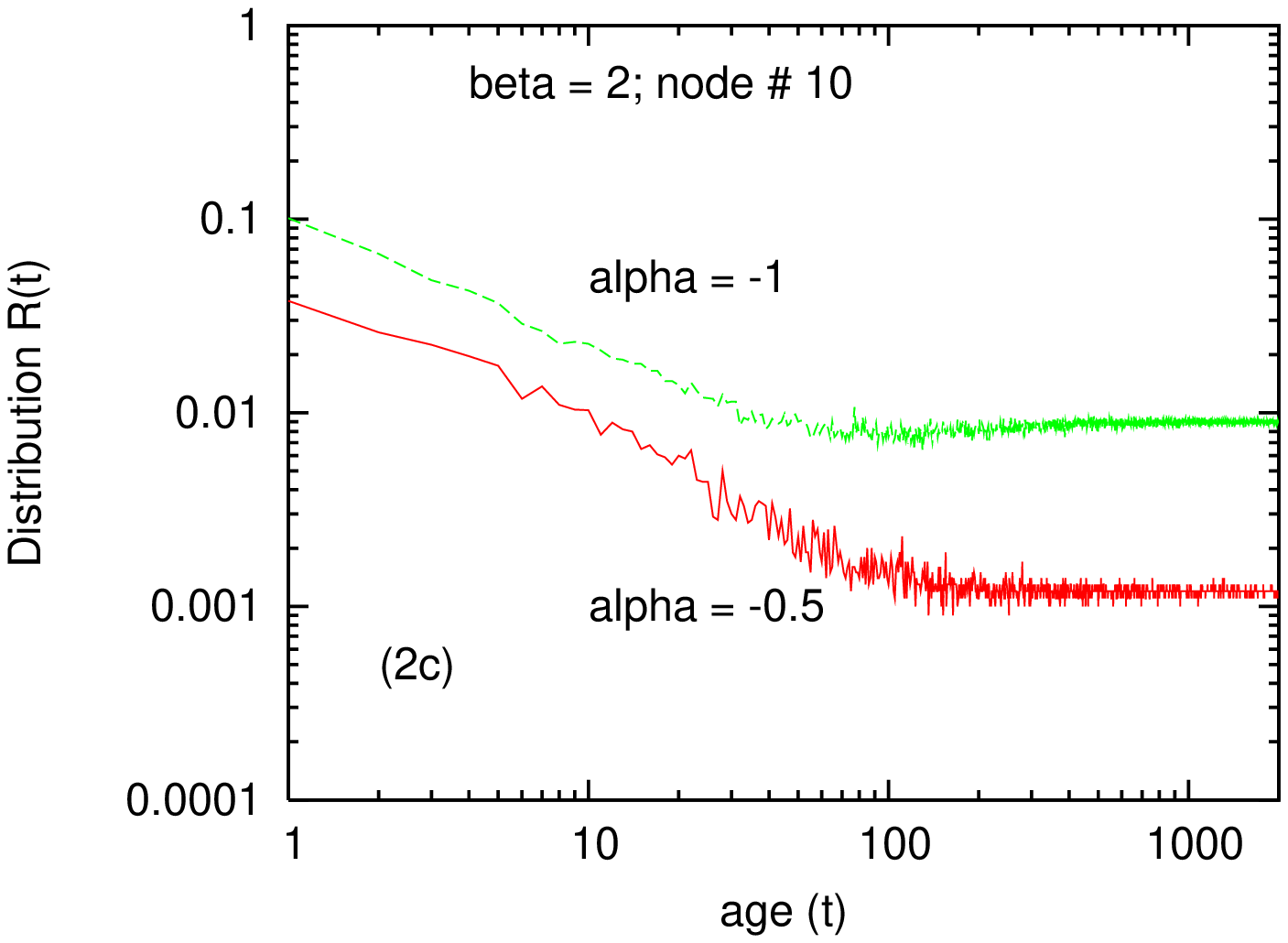}
\caption{$R(t)$ vs $t$   are shown for $\beta=0.5,1$ and $2$. In figs 2a and 2b,
 the variations are shown for $\alpha =0, -0.5, -1.0$. Here, the variation 
is power law at large values of $t$ only. 
For $\beta =2$, variations are shown for $\alpha = -0.5$ and $-1.0$. Here however, $R(t)$ behaves differently; a  
power law variation exists for early $t$ and dies out very soon to a saturation value
(fig 2c).}
\end{figure}
\end{center}

For $\beta=0.5$ once again we obtained a similar variation of $R(t)$. Power law 
regions might exist for $\alpha = -0.5$ and $-1.0$ with exponents $ \sim 0.8, 1.0$
respectively. However, $\beta < 1$ may not be a very interesting region as it
has already been found that there is no scale-free
behaviour here.

For $\beta=2$  behaviour  of $R(t)$ changes: there is apparently 
a power law region with exponent $\sim 0.7$ during early
times and later it becomes a constant. The later 
behaviour is not consistent with the citation results
where $R(t)$ decays rapidly for large $t$.
These 
results for the three different $\beta$  are shown in Fig. 2.

\medskip

III.2. {\it Scheme (b)}:
The  attachment probability for the second scheme is given by 
\begin{equation}
\Pi(k,t) \sim k^{\beta}\exp({\alpha}t).
\end{equation}
 
In \cite{red2} and \cite{zhu} the behaviour of $T(t)$ was found to be exponentially decaying. We have therefore taken a model with $f(t) = \exp({\alpha}t)$ with
$\alpha <0$. We have also generalised the model of \cite{zhu} to include
a nonlinear functional dependence of $\Pi(k,t)$ on $k$.
This is because the $\beta=1$ case showed that there is no scale free
region for negative $\alpha$. A scale free region may only be obtained for
values of $\beta > 1$ when  $\alpha < 0$. 
  
\begin{center}
\begin{figure}
\includegraphics[clip,width= 5cm,angle=270]{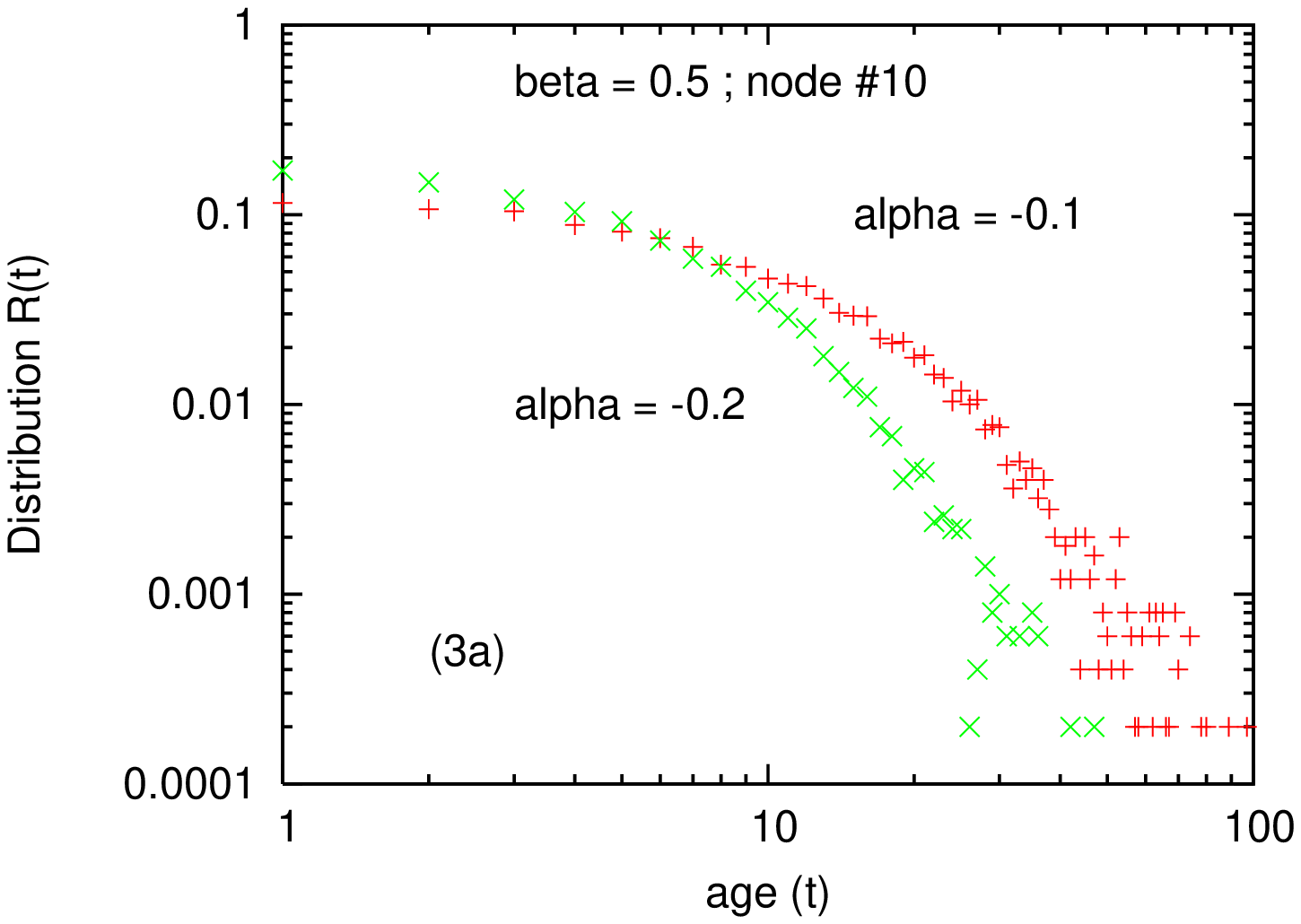}
\includegraphics[clip,width= 5cm,angle=270]{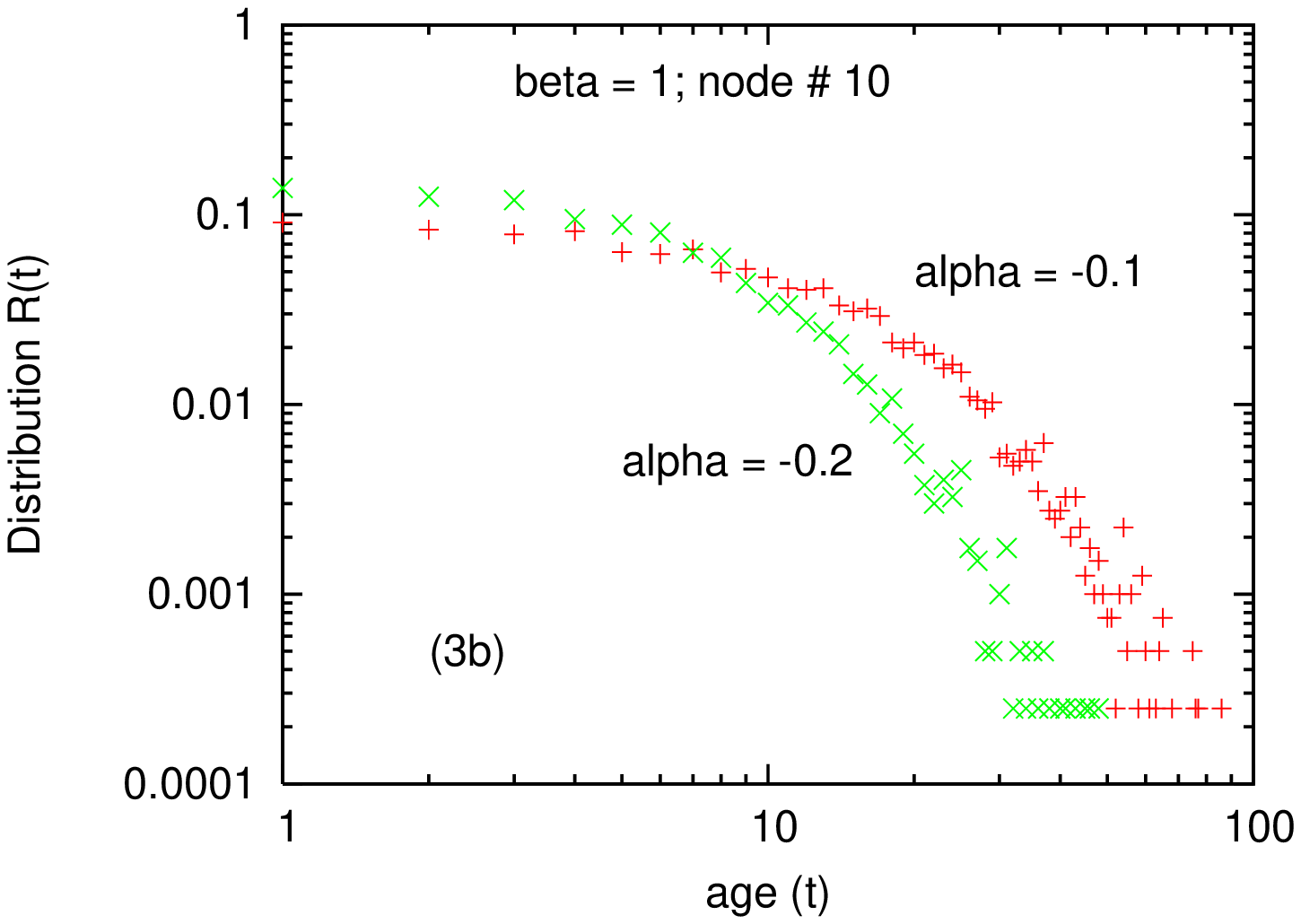}
\includegraphics[clip,width= 5cm,angle=270]{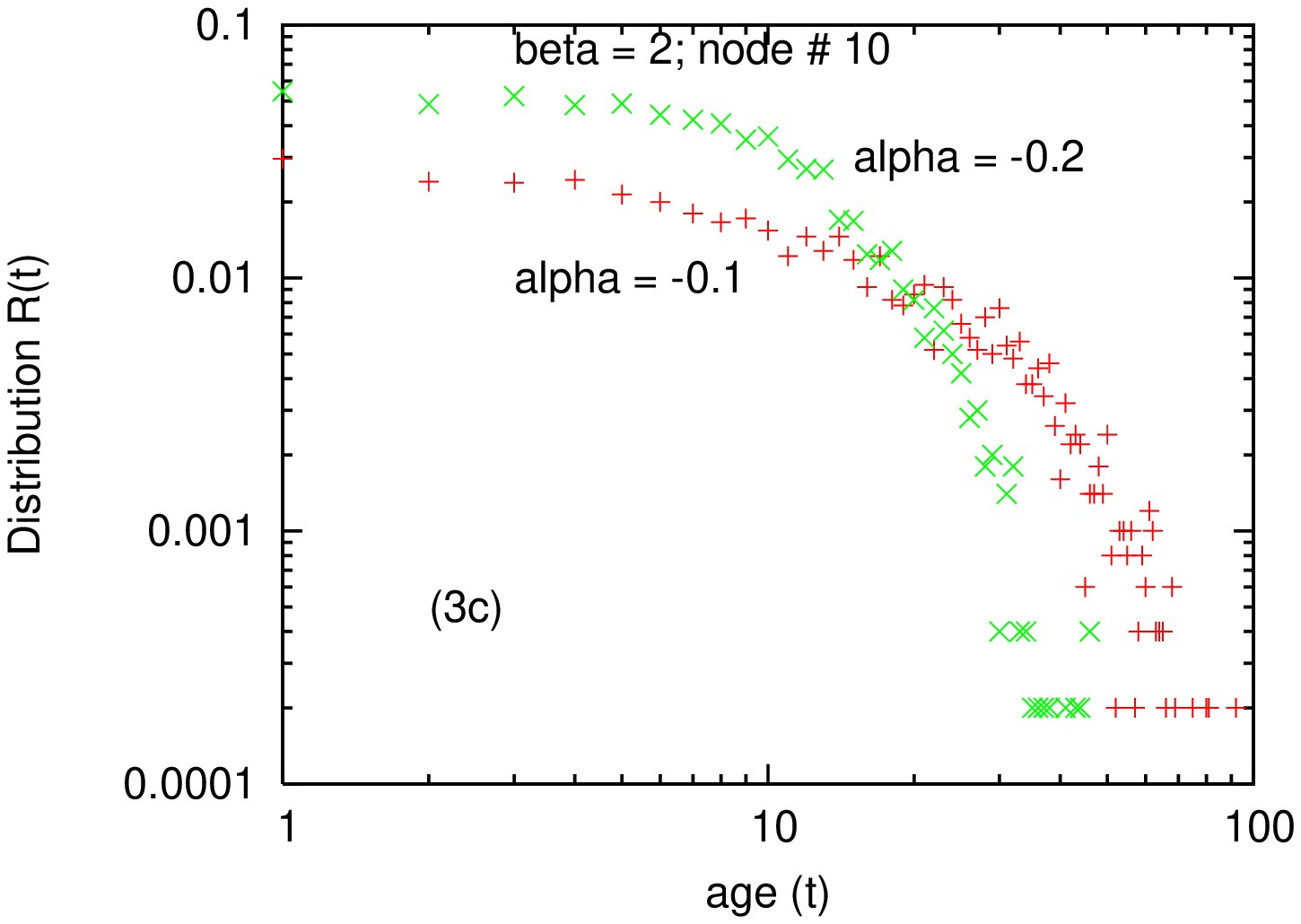}
\caption{ $R(t)$ vs $t$ data are shown at $\beta =0.5,1,2$ respectively for 
$\alpha=-0.1,-0.2$. Power law is not observed here at all.}
\end{figure}
\end{center}
For the exponential time dependence in $\Pi(k,t)$, once again we study $R(t)$,
the  age distribution of the citations to a tagged paper 
 for the values of $\beta=0.5,1.0,2.0$ keeping $\alpha \le 0$.  

In this case, power law is not obtained anywhere for $R(t)$.
For each value of $\beta$, we show  in Fig. 3
$R(t)$  for $\alpha = -0.1$ and $-0.2$ (these values are comparable to the 
observed values).

\section{A Modified Model: R(t) and other results}

We are in search of a minimal model and find that the simple models described in
the previous section are not sufficient.
To add more features, we note that there are many differences between these 
models and a real citation network, prominent among which are the following\\
(i) In these models, only one paper is being cited by each paper\\
(ii) In each year, it is being assumed in these models that 
only one paper is being published. (Note that
the unit of time for the real data had been 1 year). \\
Both these are gross simplifications and the real network is 
quite different.

In order to make the smallest changes, we incorporate suitable modifications
in the models described in section III such that only one of the two factors 
mentioned
is considered at a time. This way, it will be also 
be clear which are the indispensible features of the citation network.

We take the exponential model where the attachment probability is given by 
(6) because we wish to proceed with a model in which  the 
time dependent part in the attachment probability has an exponential
decay to mimic reality. 

Keeping everything else same, when each new node is allowed to have more 
than one citation
(typically 10 or 20) we find that there is no  
significant change in the behaviour of $R(t)$.

Next, again sticking to the exponential model with one citation, 
we consider $M$ number
of publications each year ($M > 1$).
 In the simulation, this means 
we are putting the time label differently, the first $M$ nodes have $\tau_i =1$,
the next $M$ nodes $\tau_i=2$ etc. With $M=20$,
we find that the behaviour of $R(t)$ is indeed a power law for $t \leq 20$,
when the value of $\beta=2$ and $\alpha = -0.1,-0.2$ with an
exponent $\rho = 1.4 \pm 0.1$. (Fig. 4).
Decreasing the value of $\beta$, the power law behaviour worsens.
There maybe some optimum values of $\beta$ and $\alpha$ for which the
value of the exponent $\rho$ is closer to the observed 0.94 \cite{red2} or some more
modifications
of the basic model maybe required to achieve a better quantitative 
agreement.  Our present objective is not to
obtain precise values but rather to obtain the simplest possible model
that has an exponentially decaying $f(t)$  giving a power law decay
in $R(t)$.  

Once we have achieved the primary goal, it is important to
find out the behaviour of the degree distribution $P(k)$ and 
the correlation between average age of citations $A(k)$ to a paper and
its degree $k$. 
Here we find that $A(k)$ has a power law type increase as has been
observed in \cite{red2} when $M=20$ for $\beta =2$ and $\alpha =-0.2$.
 For $M=1$, which corresponds to the model described in III.2,
 it is definitely not a power law (Fig. 6).
 Hence at least two features of the present model are 
 consistent with the observations of \cite{red2}.

Lastly, we check the degree distribution. For a few initial decades of $k$, it does 
give a fairly good agreement with a power law decay of the form 
$P(k) \sim k^{-\gamma}$ with $\gamma = 3$. However,
there is a  increase in $P(k)$ for very large $k$ values which
indicates a tendency to form a gel (Fig. 5). 
In fact, the curvature of $P(k)$ is opposite to that 
of the observed distribution reported in \cite{red1,vazquez}. The possible
 reasons for this departure from reality is discussed briefly in the next
section.

\begin{figure}
\includegraphics[clip,width= 5cm,angle=270]{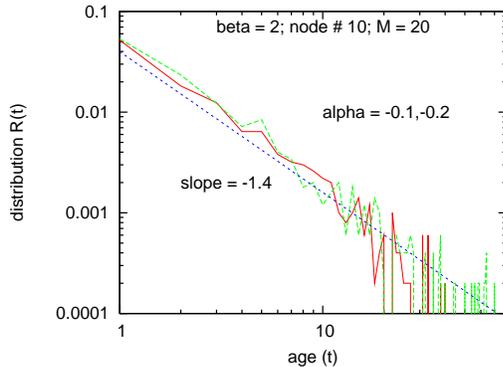}
\caption{$R(t)$ vs $t$ plot with redefined time, i.e, now $M$ nodes are born 
in each year. Here $M=20$. A power law behaviour is obtained  
for $\beta=2$ at values of $\alpha =-0.1$ (dashed line) and $-0.2$ (solid line)
 with exponent 
$\rho=1.4 \pm 0.1$.   It is observed that as  $|{\alpha}|$ increases, the 
 power law breaks down at an earlier $t$.}
\end{figure}

\begin{figure}
\includegraphics[clip,width= 5cm,angle=270]{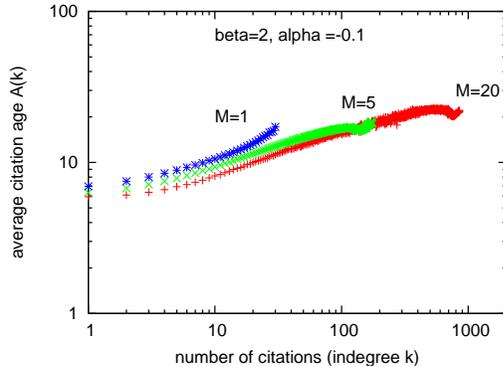}
\caption{Average citation age versus number of citations for $M = 1, 5, 20$, where $M$ is the number of nodes born per time step. Here $\beta=2.0$ and 
$\alpha = -0.2$. As expected, there is a positive correlation between $A(k)$ 
and $k$, and for larger values of $M$ it fits to a power law dependence.}
\end{figure}

\begin{figure}
\includegraphics[clip,width= 5cm,angle=270]{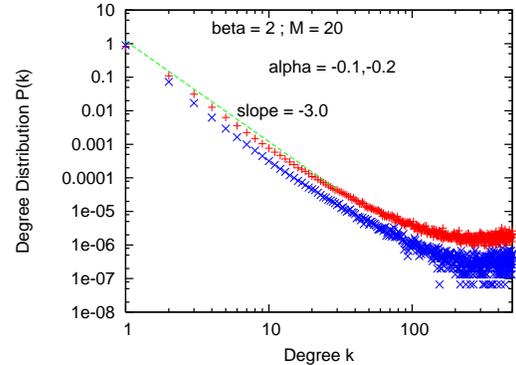}
\caption{This figure shows the degree distribution $P(k)$ 
for $\beta =2$ , $\alpha = -0.1,-0.2$ and $M=20$ .  The straight 
line  with the slope indicated shows the closest fit for $k < 100$. } 
\end{figure}

\section{Summary and Conclusion}

We have attempted to construct a simple model for citation network in which 
the evolution rule is formulated according to the behaviour of real
citation data.

Since aging is an important factor in citation data, our emphasis has been 
on the age distribution of references made {\it by} a paper $T(t)$ and made 
{\it to} a paper $R(t)$. The interesting observation was that $R(t)$ has a 
power law decay for early $t$ while $T(t)$ has an exponential decay, which 
is rather counter-intuitive. Indeed, the standard aging network models 
 fail,
 but simple modification of the exponential model is able to reproduce the
correct behaviour of $R(t)$, at least qualitatively.

It is in general not quite easy to construct a single model of citation 
network which can reproduce all its features \cite{vazquez,ps2}. This may
 be due to certain distinctive features of the citation network of which
we mention a few below.

(i) Apart from mathematical quantities like the degree and age of a paper, the
content of a paper is also important. Evidently a paper on a topic where a
 large number of people work, will get more citations ( that can be quantified 
by the {\it impact parameter} of a paper). \\
(ii) Neither the number of citations nor the number of papers published each 
year remains constant. \\
(iii) In the models, one assumes smooth behaviour, e.g., of $T(t)$ while
in reality the variations  are non-monotonic.
(iv) There is a  possibility of "death" of a papers, or 
the separate existence of dead and live papers as referred to in \cite{sune}. 

In our modified model, although we have obtained good agreement of the 
behaviour of $R(t)$ and $A(k)$, but  for $P(k)$
the behaviour does not agree very well with the observations.
This may be because we have not optimised the values
of $\beta$ and $\alpha$ to get better agreement with the 
real data and also due to the reasons stated above.\\

 Acknowledgments: KBH is grateful to CSIR (India) F.NO.9/28(609)/2003-EMR-I for financial support.
PS acknowledges CSIR grant no.  03(1029)/05/EMR-II.

Email: kamalikabasu2000@yahoo.com, psphy@caluniv.ac.in

\end{multicols}

\begin{thebibliography}{99}
\bibitem{explain} In \cite{kamo1.ps} a power law variation of $T(t)$ was claimed for a small sample and using a normalisation procedure; however, for
observational data such normalisation is not meaningful.
\end{thebibliography}
\end{document}